\documentclass[11pt]{article}
\usepackage[hyperref]{acl}
\usepackage{pgfplotstable}
\usepackage{times}
\usepackage{latexsym}
\usepackage{microtype}
\usepackage{amsmath}
\usepackage{amssymb}
\usepackage{amsthm}
\usepackage{booktabs}
\usepackage{multirow}
\usepackage{graphicx}
\usepackage{xcolor}
\usepackage{subcaption}
\usepackage{enumitem}
\usepackage{float}
\usepackage{url}
\usepackage{tikz}
\usepackage{pgfplots}
\pgfplotsset{compat=1.18}
\usetikzlibrary{patterns}
\usepackage{hyperref}
\hypersetup{
  pdftitle={Per-Query Gating of LLM Rerankers for Multi-Hop Retrieval},
  pdfauthor={Andre Bacellar},
  pdfkeywords={LLM reranking, per-query routing, multi-hop retrieval, HippoRAG,
               calibrated fusion, conformal prediction, cost-quality Pareto},
}

\newcommand{\system}{\textsc{GateRR}}

\newcommand{\hipporagtwo}{\textsc{HippoRAG2}}
\newcommand{\phasegraph}{\textsc{PhaseGraph}}
\newcommand{\bridgerag}{\textsc{BridgeRAG}}
\newcommand{\regimerouter}{\textsc{RegimeRouter}}
\newcommand{\proprag}{\textsc{PropRAG}}

\newcommand{\LH}{\mathrm{LH}}

\newcommand{\rerank}{\mathrm{rerank}}

\title{Per-Query Gating of LLM Rerankers for Multi-Hop Retrieval}

\author{Andre Bacellar\thanks{Independent researcher.} \\
  \href{mailto:andremi@gmail.com}{\texttt{andremi@gmail.com}}}

\begin{document}
\maketitle

\begin{abstract}
LLM rerankers add of the order of \$0.2--0.3 per 1{,}000 queries and about
a second of tail latency on top of a graph-augmented dense pipeline such as
\hipporagtwo{}, and on our three multi-hop benchmarks they improve the
final-hop passage's top-$K$ coverage on seven of nine (dataset, $K$) cells,
by up to $+34.8$\,pp, while two cells are saturated.
We ask whether a learned per-query gate can skip the reranker where it will
not help, using only features available before the LLM call: 27 score and
lexical statistics of the two retrieval lists plus a PCA of a small query
embedding, with an executable fallback (\hipporagtwo{}'s own top-$K$, or the
top-$K$ of a distilled per-passage scorer trained inside the fold).
We evaluate under a nested protocol in which every choice, including the
fallback and the threshold, is made inside the training fold and applied
once to held-out queries, and we report harmful skips (the rerank would have
found the target, the fallback did not) next to the aggregate coverage.
Across nine cells on 2WikiMultiHopQA, MuSiQue and HotpotQA the gate skips
$51\%$ of reranker calls on average at an average held-out LastHop@$K$ cost
of $1.2$\,pp.
Four cells meet a pre-registered non-inferiority rule of $1$\,pp (skipping
$36.5$, $45.1$, $97.7$ and $99.6\%$ of calls); the other five buy $21$ to
$54\%$ savings at $1.2$ to $3.3$\,pp, and harmful skips occur in eight of
nine cells ($190$ harmful against $136$ beneficial over all cells).
A random gate at the same skip rate loses $2$ to $11$\,pp on the high-lift
cells, so the decisions carry signal, but they are not lossless.
A second rule sets each cell's threshold from a pre-specified budget on the
expected harmful-skip rate over Platt-calibrated harm probabilities (ECE
$0.025$ after calibration, $0.094$ before).
At a $1$\,pp budget the gate skips $42\%$ of calls at $-0.8$\,pp with $66$
harmful skips and six cells within $1$\,pp, but the realised harm exceeds
the promised harm in six of nine cells (mean $1.45$ against $0.83$\,pp), a
selection optimism we quantify; a $0.5$\,pp budget realises about $1$\,pp
($33\%$ skipped, $-0.5$\,pp, seven cells within $1$\,pp).
The harm probabilities are calibrated but barely discriminative (AUC $0.16$
to $0.70$): the savings come from cells where harm is rare, not from telling
harmed queries apart.
An earlier version of this paper reported $73\%$ average ``lossless''
savings; that figure rested on a fallback that used the gold label and on a
wrong MuSiQue target, and we document both.
\end{abstract}

\section{Introduction}
\label{sec:intro}

Graph-augmented dense retrieval --- \hipporagtwo{}~\cite{gutierrez2025hipporag2},
\proprag{}~\cite{proprag2025} and calibrated graph-vector fusion
(\phasegraph{})~\cite{phasegraph2026} --- has become the standard recipe for
multi-hop QA.
A further boost is available by appending a large-language-model reranker to
the merged candidate list~\cite{llama3_2024,sun2023rankgpt}.
On our 2Wiki evaluation set ($n{=}491$), \hipporagtwo{} alone puts the
final-hop passage in its top-5 for $62.3\%$ of queries, and the same
candidates reranked by Llama-3.3-70B for $80.0\%$ ($+17.7$\,pp, $W/L{=}87/0$).
The reranker wins at every cutoff on 2Wiki and MuSiQue and at $K{=}1$ on
HotpotQA (Table~\ref{tab:main-lift}); on HotpotQA at $K{=}5$ and $10$ the
free retrieval is already saturated and the reranker changes nothing.

\paragraph{The cost problem.}
That lift is paid for on every query.
A Llama-3.3-70B rerank of a merged top-20 list consumes $1{,}007$ to
$1{,}367$ prompt tokens and $27$ to $36$ completion tokens per query, which at
the Nebius prices of the time (\$0.20 per million prompt, \$0.60 per million
completion tokens) is \$0.22 to \$0.29 per 1{,}000 queries in marginal
reranker cost, and a network call of the order of a second at the tail.
Today's deployed graph-RAG stacks pay both on every query whether or not the
rerank will change the answer.

\paragraph{The opportunity.}
The lift is not uniform.
On every cell there are queries where \hipporagtwo{}'s own top-$K$, or the
top-$K$ of a cheap distilled scorer, already holds the target, and an oracle
that skipped exactly those queries would save $55$ to $100\%$ of calls
without harming a single query (Table~\ref{tab:baselines}, last column).
The question is how much of that an actual gate, trained on features
available before the LLM call, can capture on queries it has never seen.

\paragraph{What this version changes.}
An earlier version of this paper reported $73\%$ average ``lossless''
savings and a live deployment.
An independent review found that three of the nine cells scored a fallback
that took the per-query maximum of two hit indicators, which needs the gold
label and is therefore an oracle; that the ``held-out'' audit evaluated the
classifiers on their own fitting queries; that the reported safe-skip rate
was skip rate plus the negative part of the net coverage change rather than
a count of harmed queries; and that the MuSiQue target was the last
supporting paragraph in the dataset's shuffled paragraph order, the terminal
hop for only $45\%$ of test queries, so that a reported $10$\,pp reranker
loss on MuSiQue $K{=}1$ was in fact a $10.9$\,pp win.
This version fixes all four: executable fallbacks only, a nested evaluation
in which every choice is made inside the training fold, per-query harm
accounting with a pre-registered non-inferiority rule, and corrected targets
on MuSiQue and HotpotQA.
The numbers are smaller and, we believe, right.
The second arXiv version adds the loss-budget rule of
Section~\ref{sec:method:budget}: Platt-calibrated harm probabilities, a
threshold computed inside the training fold from a stated expected-harm
budget, and the realised harm reported against the promised one
(Section~\ref{sec:exp:budget}); the deployed policy is retrained under that
protocol (Section~\ref{sec:production}).
No number of the first version changed.

\paragraph{Contribution.}
We introduce \system{}, a per-query learned gate over a two-source pipeline
(\hipporagtwo{} $\oplus$ \phasegraph{}) reranked by an LLM.
\begin{itemize}[noitemsep,leftmargin=*]
\item A gate whose 27 hand-crafted features are computed by the same function
      in research and in serving, optionally extended with a PCA of a
      \textsc{BGE-small} query embedding fitted inside the fold, and two
      executable skip-fallbacks: \hipporagtwo{}'s top-$K$ and the top-$K$ of a
      per-passage distilled scorer trained inside the fold on cached LLM
      relevance scores.
\item A nested evaluation protocol (Section~\ref{sec:method:nested}) and a
      reporting standard that puts harmful and beneficial skips next to the
      aggregate coverage change, with baselines a practitioner would try
      first: a one-feature score-gap gate, the best fixed action per cell, and
      a random gate at matched cost.
\item Held-out results on nine (dataset, $K$) cells (Table~\ref{tab:gate-heldout}):
      $51\%$ of reranker calls skipped on average at an average cost of
      $1.2$\,pp LastHop@$K$; four cells non-inferior at $1$\,pp, five not;
      harmful skips in eight cells; the learned gate well ahead of a random
      gate at the same cost on every high-lift cell and no better than a
      fixed action on the saturated ones.
\item Corrected targets for MuSiQue (terminal decomposition passage) and
      HotpotQA (answer-bearing supporting passage), with the reranker's lift
      re-measured against them (Table~\ref{tab:main-lift}).
\item A loss-budget selection rule (Section~\ref{sec:method:budget}) whose
      threshold is computed inside the training fold from a pre-specified
      expected-harm budget over Platt-calibrated harm probabilities, scored
      on the same held-out folds with expected and realised harm side by
      side (Section~\ref{sec:exp:budget}): $42\%$ of calls skipped at
      $-0.8$\,pp under a $1$\,pp budget, six cells within $1$\,pp, realised
      harm above the promise in six cells, calibration good in aggregate
      (ECE $0.025$) and the harm ranking weak (AUC $0.16$ to $0.70$).
\end{itemize}

\section{Related Work}
\label{sec:related}

\paragraph{Graph-augmented dense retrieval.}
Graph-based retrieval enriches an underlying dense passage index with structural
signals (entity co-occurrence, propositional links, learned bridges).
\hipporagtwo{}~\cite{gutierrez2025hipporag2}, the successor of
HippoRAG~\cite{gutierrez2024hipporag}, couples a passage-graph
personalized-PageRank (PPR) walk with dense ANN retrieval over an
\textsc{NV-Embed-v2}~\cite{nvembed2024} index: PPR mass concentrates on passages
that are graph-neighbours of dense-similar seeds, capturing the multi-hop
structure of questions that single-vector ANN misses.
\proprag{}~\cite{proprag2025} replaces PPR with beam search over proposition
paths, trading per-query latency for explicit-chain interpretability.
Our earlier \phasegraph{} paper~\cite{phasegraph2026} introduces a
PIT-Boltzmann score calibration that lets graph and vector scores be linearly
mixed at a per-dataset thermo $\alpha$.
All three families are \emph{always-on}: every query pays the graph traversal
cost, regardless of whether the question is multi-hop enough to benefit.
Our work treats one such pipeline (\hipporagtwo{} $\oplus$ \phasegraph{}) as the
free retrieval substrate and asks the orthogonal question of when an even more
expensive layer (the LLM reranker) should be invoked.

\paragraph{Calibrated fusion and ensemble retrieval.}
A long line of work has explored how to combine multiple retrievers without
re-training.
Reciprocal Rank Fusion (RRF)~\cite{cormack2009reciprocal} averages the
reciprocal ranks of items across multiple lists and has remained surprisingly
competitive because it is parameter-free and rank-invariant under score scaling.
Linear interpolation of $\mathrm{BM25}$ and dense scores requires per-corpus
score calibration~\cite{guo2017calibration} but is the dominant approach in
modern hybrid retrieval; \phasegraph{}'s contribution is to extend this
calibration to graph-vector mixtures via per-source temperature scaling
followed by Boltzmann normalization.
Across all these methods, ensembling is a coverage-preserving operation:
multiple retrievers' candidates union into a richer pool but no \emph{strict
ordering} information beyond rank is introduced.
LLM reranking changes that: it reorders the merged list using semantic
reasoning that the score-level fusions cannot replicate, and our Phase-2
ablations (Section~\ref{sec:exp:main}) show 8 score-only fusion variants all
reaching K=5 coverage $\leq 0.623$ on 2Wiki vs.\ LLM rerank's $0.800$
(a $\geq 17$\,pp gap).
The LLM rerank is therefore a fundamentally different operation than fusion,
and per-query gating is the right level at which to manage its cost.

\paragraph{LLM reranking for retrieval.}
A growing body of work appends a large generative LM to rescore a small
candidate set returned by a cheap first-stage retriever.
\textsc{RankGPT}~\cite{sun2023rankgpt} demonstrated that frontier LLMs can be
prompted listwise (``rank these passages 1\ldots20'') and approach or beat
supervised cross-encoders.
\textsc{RankZephyr}~\cite{pradeep2023rankzephyr} distills these listwise
rankings into a smaller open model for cost-efficient deployment.
A parallel line of work uses supervised rerankers trained on relevance
pairs: cross-encoders such as \textsc{monoBERT}~\cite{nogueira2019passage} and
\textsc{BGE-reranker}~\cite{xiao2023bge}, and the late-interaction model
\textsc{ColBERT}~\cite{khattab2020colbert}.
Our Phase-2 baseline shows that \textsc{BGE-reranker-large}, the strongest open
cross-encoder we tested, reaches only $0.595$ at $K\!=\!5$ on 2Wiki vs.\
Llama-3.3-70B's $0.800$ (a $-20.6$\,pp gap) at roughly $50\times$ lower
inference cost.
This is the cost-quality tradeoff that motivates per-query gating: the LLM is
\emph{both} more accurate \emph{and} an order of magnitude more expensive than
its cheaper alternatives, so the policy question is when to spend the cost.
We treat the LLM reranker as a fixed ceiling module and ask only \emph{when}
to invoke it.

\paragraph{Learned routing for retrieval-augmented generation.}
Several recent systems route between retrieval modes.
Self-RAG~\cite{asai2024self} trains a critic to decide whether retrieval is
useful per-token; Adaptive-RAG~\cite{jeong2024adaptive} classifies questions
into single-hop / multi-hop complexity classes and routes to different
retrievers per class; Self-Route~\cite{wang2024selfroute} chooses between
short-context retrieval and long-context direct prompting.
The shared intuition is that the cost of retrieval (or the next pipeline
stage) is non-uniform across queries, so a learned router can save inference
without losing accuracy.
None of these papers routes the LLM reranker specifically; the
gate-then-rerank decision sits at a coarser cost-quality knee.
In our own family of work, \regimerouter{}~\cite{bacellar2026regime} routes
between query-only retrieval and graph-fusion via a regime-conditional
classifier and matches always-fusion coverage at $52\text{--}90\%$ lower
compute; we adapt the same per-query routing intuition to the more expensive
LLM-rerank choice, selecting the gate under a coverage constraint against
always-rerank rather than against the cheap baseline, and reporting where
that constraint fails to hold out of sample.
\bridgerag{}~\cite{bacellar2026bridgerag} uses an LLM judge $s(q,b,c)$ at
query time to score bridge passages, which is the \emph{opposite} cost
profile to ours: \bridgerag{} adds query-time LLM cost on top of dense
retrieval, while \system{} amortizes it away on the cells where it would not
have helped.
\regimerouter{}'s companion, RegimeAbstain~\cite{bacellar2026abstain},
predicts \emph{retrieval failure} from the same kind of score-distribution
features we use here; a calibrated failure probability is a natural upstream
signal for a gate, which we leave to future work.

\paragraph{Concurrent work on adaptive and cost-aware RAG (2025--2026).}
Several papers that appeared while this work was in preparation route on
cost.
Retrieval-as-a-Decision~\cite{wang2025retrievaldecision} gates
\emph{whether to retrieve at all} without training; RAGRouter-Bench
\cite{bansal2026ragrouter} benchmarks lightweight routing across retrieval
strategies of different token cost; CA-RAG~\cite{mishra2026costaware} picks a
retrieval \emph{depth} per query by maximizing a utility over quality,
latency and token cost; Adaptive Re-Ranking~\cite{genc2026adaptiverr}
adapts the re-ranking stage itself; and BalanceRAG~\cite{jia2026balancerag}
certifies threshold pairs for an LLM-only$\to$RAG cascade at a prescribed
risk level using sequential graphical testing.
Two works are closer still.
ARDS~\cite{gonzalez2026ards} trains a lightweight classifier on retrieval
signals to decide per query whether a cross-encoder reranker should run at
all, the same rerank/no-rerank framing as ours with a cheaper reranker;
\citet{dabod2026rerankinghurts} gate a few-shot reranker with an
uncertainty estimate at a calibrated operating point and document the cases
in which reranking degrades results.
These share our premise that stage cost is non-uniform across queries.
They differ in the routed object: none gates a listwise LLM reranker over a
merged graph-vector candidate set with an executable distilled fallback, and
none reports per-query harm alongside the aggregate coverage constraint.
We do not claim the framing itself as new.
BalanceRAG's risk control is complementary rather than competing: our
per-cell threshold $T$ is selected by out-of-fold sweep, and the $3$--$7.5$\,pp
held-out regressions we report in Section~\ref{sec:production} are exactly
the kind of gap a certified threshold procedure is designed to close.

\paragraph{Coverage-versus-$K$ analyses.}
Conformal prediction~\cite{vovk2005algorithmic,angelopoulos2021gentle}
turns a scoring rule into prediction sets with a distribution-free coverage
guarantee, given a calibration sample that is disjoint from the test sample
and a fixed, nested family of sets.
An earlier draft of this paper reported ``conformal prediction-set sizes''
for the gated pipeline; those numbers were computed on the same queries used
to choose the threshold and with missing cutoffs filled from larger ones, so
they were neither conformal nor exact.
Section~\ref{sec:exp:coverage} now reports plain empirical coverage-versus-$K$
curves on held-out queries for the always-rerank pipeline and for the gated
pipeline, with no guarantee claimed.

\paragraph{Distillation of LLM rerankers.}
LLM relevance scores can be distilled into cheaper scorers.
\textsc{RankZephyr}~\cite{pradeep2023rankzephyr} distills frontier-LLM
listwise rankings into a 7B model; classical cross-encoder distillation
(e.g.\ from cross-encoder to bi-encoder) underlies most of the dense-retrieval
pipeline.
Our distillation is a per-(query, passage) GBM regression on the cached
Llama-3.3-70B relevance scores of the merged candidates, with features derived
from \textsc{BGE-small} embeddings and rank positions in the upstream
retrievers.
We do not use the distilled scorer as a standalone retriever; it is one of the
two executable \emph{skip-fallbacks} the gate can choose (the other is
\hipporagtwo{}'s own top-$K$), and it is trained inside every evaluation fold
(Section~\ref{sec:method:fallback}).

\section{Method}
\label{sec:method}

\subsection{Two-source retrieval}
\label{sec:method:retrieval}
For each query $q$ we compute two candidate lists in parallel.
The first is $H(q)$: \hipporagtwo{} top-10 with PPR scores, using
\textsc{NV-Embed-v2}~\cite{nvembed2024} as the dense edge-weight encoder over
the official \hipporagtwo{} corpus index (6{,}119 unique titles on 2Wiki, 1{,}099
on MuSiQue, 9{,}811 on HotpotQA).
The second is $P(q)$: \phasegraph{} calibrated fusion top-10 with the
${\mathrm{thermo}}\_\alpha^\star_{d_k=30}$ config that maximizes per-dataset
LH@10 in a held-out probe (2Wiki: $\alpha\!=\!0.4,\,d_k\!=\!20$; MuSiQue:
$\alpha\!=\!0.7,\,d_k\!=\!30$; HotpotQA: $\alpha\!=\!0.5,\,d_k\!=\!50$).
The two sources are computed without any LLM in the loop; their total
amortized latency in production is dominated by the dense ANN query
($\approx 50$\,ms per source on a CPU-backed FAISS index), and the candidate
overlap between $H(q)$ and $P(q)$ in their respective top-10 averages 4--6 of
10 across our three datasets.
The two-source split is therefore both cheap and complementary: it expands
the candidate pool the LLM rerank can choose from, and the per-source score
statistics constitute most of the features used by the gate
(Section~\ref{sec:method:gateV5}).

\subsection{LLM rerank}
\label{sec:method:rerank}
We merge $H(q)\cup P(q)$ to a unique top-20 set $M(q)$ (de-duplicated by
title), then call Llama-3.3-70B-Instruct via Nebius with one of two prompts.
\emph{Prompt v1} is relevance-centric: ``Given this question, score each
passage by relevance (0--10) for finding the answer. Output ONLY a
comma-separated list of $|M|$ integers.''
\emph{Prompt v2} is answer-centric and adds two sentences: ``Exactly one
passage contains the answer; bridges (intermediate context) get $\leq 7$;
only the passage that contains the answer gets 10.''
The model is invoked at temperature 0 with at most 80 completion tokens;
candidates are reordered by predicted score (ties broken by original merged
rank) and the top-$K$ list is returned.
An answer-centric prompt variant (``prompt v2'') was also cached on MuSiQue
during the earlier draft; its comparison rested on the wrong MuSiQue target
and it is not used anywhere in this version, which scores prompt v1
throughout.

\subsection{Learned gate}
\label{sec:method:gateV5}
The gate predicts, before the LLM call, whether the rerank's top-$K$ will
improve on a cheap fallback's top-$K$ for this query.
We train one classifier per (dataset, $K$) cell.
Every number reported for the gate in Section~\ref{sec:exp} comes from a
nested cross-validation in which the classifier, its features' PCA, the
label, the threshold and the fallback are all chosen inside the training
fold and applied once to queries the selection never saw
(Section~\ref{sec:method:nested}).

\paragraph{Features (27 hand-crafted $+$ PCA).}
The 27 hand-crafted features are computed by the same function that serves
the gate in production (\texttt{build\_v2\_features}); Table~\ref{tab:features}
lists them in code order.
Optionally, a PCA projection of the \textsc{BGE-small} query embedding
(8, 16 or 32 components, fitted on the training fold only) is appended.
\begin{table*}[t]
\centering
\small
\begin{tabular}{l c p{0.64\textwidth}}
\toprule
Block & Count & Features (in code order) \\
\midrule
\hipporagtwo{} scores $s_H$ & 9 & $s_H[1], s_H[2], s_H[3], s_H[5]$; $s_H[1]\!-\!s_H[2]$; $s_H[1]\!-\!s_H[5]$; $\max$, mean, std over the top-10 \\
Vector scores $s_V$ & 7 & $s_V[1], s_V[2], s_V[5]$; $s_V[1]\!-\!s_V[2]$; $\max$, mean, std over the top-10 \\
Cross-list agreement & 5 & Jaccard of the top-1, top-3, top-5 and top-10 title sets of $H$ and $V$; rank of $H$'s top-1 title in $V$ (10 if absent) \\
Query & 6 & characters$/200$; words$/30$; count of WH-words; comparison-pattern flag; aggregation-pattern flag; unique titles in $H$'s top-10 $/10$ \\
\bottomrule
\end{tabular}
\caption{The 27 hand-crafted gate features, as computed by the production
function \texttt{mdma\_api.paper14\_gate.build\_v2\_features}. $s_H[i]$ and
$s_V[i]$ are the $i$-th ranked \hipporagtwo{} and dense scores (0 if the list
is shorter). Missing values are 0. The same function produces the research
feature matrix, so the research and serving schemas are identical.}
\label{tab:features}
\end{table*}

\paragraph{Labels.}
Both labels are defined relative to the fallback $F\in\{H, D\}$ the cell
uses (Section~\ref{sec:method:fallback}).
\textsc{rerank\_helps} is $\mathbb{1}[\LH^{\rerank}_K(q)>\LH^{F}_K(q)]$, ``does
the rerank strictly beat the fallback on this query at this $K$''; it is
sparse wherever the fallback is already strong.
\textsc{fallback\_has\_gold} is $\mathbb{1}[\text{gold}\in F_K(q)]$; it is
better balanced, and predicting it well is sufficient for a gate: if the
fallback already holds the gold, skipping the rerank cannot lose it.
The label is chosen per cell inside the training fold.

\paragraph{Nested selection.}
\label{sec:method:nested}
For each dataset the evaluation queries are split into five outer folds
(shuffled, fixed seed).
For each outer fold, everything is fitted on the four training folds only:
the distillation regressor behind $D$ (with inner 5-fold cross-fitting so
that the training queries' $D$ rankings are themselves out-of-fold), the PCA
of the query embedding, and one classifier per candidate configuration in
$\{H, D\}\times\{\textsc{rerank\_helps}, \textsc{fallback\_has\_gold}\}\times
\{\text{LR}, \text{GBM}, \text{RF}\}\times\{0, 8, 16, 32\}$ PCA components,
each scored by inner 5-fold out-of-fold probabilities.
For every configuration the threshold $T$ is swept in $0.02$ steps and the
policy with the largest inner savings whose inner coverage is at least the
inner always-rerank coverage is retained; the single best configuration is
then refitted on the whole training fold and applied once to the held-out
fold.
Pooling the five held-out folds gives one decision per query that no
selection step has seen.
We report, per cell, the skip rate, the pooled held-out coverage against
always-rerank, and the number of \emph{harmful} skips (rerank would have hit,
the fallback missed) and \emph{beneficial} skips (the reverse), because an
aggregate coverage constraint can be met while individual queries are
harmed.
The reporting rule was fixed before the results were computed: a cell counts
as non-inferior only if its pooled held-out coverage is within $1.0$\,pp of
always-rerank.

\paragraph{Loss-budget selection with calibrated harm probabilities.}
\label{sec:method:budget}
The coverage-constrained rule chooses the threshold that maximises inner
savings under an aggregate constraint; it has no notion of how many queries
it is allowed to harm, and its threshold is a by-product of a sweep.
This version adds a second selection rule, evaluated inside the same folds.
For every candidate (fallback, model, PCA width) we model the \emph{harm}
label $\mathbb{1}[\LH^{\rerank}_K(q)>\LH^{F}_K(q)]$, ``the query is lost if
skipped'', obtain inner out-of-fold probabilities, and map them through a
Platt calibrator~\cite{platt1999} fitted on those same out-of-fold
predictions, which removes the distortion that class re-weighting leaves in
the raw scores.
Given a budget $\beta$ on the expected harmful-skip rate, the skip set is
the largest set $S$ of training queries, taken in increasing calibrated
$\hat p(\text{harm})$, whose expected harm $\frac{1}{n}\sum_{q\in S}\hat p_q$
is within $\beta$; the threshold is the largest $\hat p$ admitted, and
beneficial skips are not credited in advance.
The candidate with the largest inner skip fraction under the budget is
refitted on the training fold and applied once to the held-out fold, where
we report the realised harm next to the expected harm the probabilities
promised, together with the calibration of $\hat p$ itself: expected
calibration error over ten equal-width bins~\cite{guo2017calibration},
Brier score and a reliability diagram.
The primary budget is $\beta=1.0$\,pp, the tolerance pre-registered for the
reporting rule; $0.5$ and $2.0$\,pp are reported as sensitivity.
This rule was added after the coverage-rule results were known; it is a
second protocol scored on the same held-out folds, not a replacement chosen
among many.

\subsection{Executable skip-fallbacks: $H$ and $D$}
\label{sec:method:fallback}
\label{sec:method:gateV7}
When the gate skips the LLM, the query is served from a fallback ranking that
must be computable without the gold label.
We use two.
$H(q)$ is \hipporagtwo{}'s own top-$K$.
$D(q)$ is the top-$K$ of a \emph{distilled} ranking: a per-(query, passage)
gradient-boosted regressor predicts the cached Llama-3.3-70B relevance score
of each candidate in the merged list $M(q)$ from 33 cheap features, exactly
those the production code computes: the cosine between the \textsc{BGE-small}
query and passage embeddings; the passage's 0-based rank in $H(q)$ and in
$V(q)$ ($-1$ if absent) and a constant placeholder for the \phasegraph{} rank;
its position in $M(q)$; the first 8 dimensions of the elementwise product and
of the difference of the two embeddings; and the first 12 dimensions of the
query embedding.
The regressor is trained inside every evaluation fold on the training
queries' cached scores (Section~\ref{sec:method:nested}); the candidates are
then sorted by predicted score.
An earlier draft of this paper scored a third fallback, the per-query maximum
of $H$'s and $D$'s hit indicators.
That quantity needs the gold label and is therefore an oracle, not a policy;
the served implementation, which placed $H$'s top-$K$ before $D$'s in a
de-duplicating merge, returned exactly $H(q)$ on every evaluated query.
The three cells whose savings rested on that fallback were the ones that
carried the earlier headline, and they are gone from this version.

\subsection{Per-cell policy}
\label{sec:method:policy}
A cell's policy is the tuple (fallback, label, model, PCA width, threshold)
selected inside the training fold; at inference it is one classifier, one
threshold and an optional projection matrix, dispatched by a dictionary
lookup on (dataset, $K$).
Under the loss-budget rule the tuple gains the Platt pair and the budget,
and the label is always the harm label; the deployed policy of
Section~\ref{sec:production} is this tuple fitted on all queries.
The policy file that was deployed during the earlier draft
(\texttt{paper14/production\_models/policy.json}) was fitted on all
evaluation queries with the oracle fallback on three cells and the wrong
MuSiQue target on three others; it is withdrawn and the deployment is
retrained from the protocol above (Section~\ref{sec:production}).

\section{Experiments}
\label{sec:exp}

\subsection{Setup}
\label{sec:exp:setup}

\paragraph{Datasets.}
We evaluate on the three multi-hop QA benchmarks reported by
\hipporagtwo{}, using the query and corpus files of its release
(Table~\ref{tab:datasets}).
\textbf{2WikiMultiHopQA}~\cite{ho2020wiki2} is constructed from Wikipedia
abstracts with explicit reasoning chains and mixes compositional,
comparison, inference and bridge-comparison questions.
\textbf{MuSiQue}~\cite{trivedi2022musique} composes single-hop questions into
2-, 3- and 4-hop ones with an explicit decomposition, whose terminal step we
use as the target.
\textbf{HotpotQA}~\cite{yang2018hotpotqa} is a bridge-and-comparison dataset
over Wikipedia paragraphs in the distractor setting.
For each dataset the evaluation queries are the half of the 1{,}000-query
file selected by an md5 hash of the query id, the same held-out halves used
in \phasegraph{}; the other half was used there for tuning and never enters
this paper.
The target passage differs by dataset and is defined in
Section~\ref{sec:discussion} (``MuSiQue and HotpotQA targets''): the terminal
decomposition passage on MuSiQue, the unique answer-bearing supporting
passage on HotpotQA where one exists, and the last listed supporting fact
otherwise and on 2Wiki.
All coverage is at the title level on the \hipporagtwo{} side because its
released outputs carry titles only.
\begin{table}[h]
\centering
\small
\resizebox{0.9\columnwidth}{!}{%
\begin{tabular}{l r r r}
\toprule
Dataset & $n$ & Records & Titles \\
\midrule
2Wiki~\cite{ho2020wiki2} & 491 & 6{,}119 & 6{,}119 \\
MuSiQue~\cite{trivedi2022musique} & 514 & 11{,}656 & 9{,}838 \\
HotpotQA~\cite{yang2018hotpotqa}  & 521 & 9{,}811 & 9{,}811 \\
\bottomrule
\end{tabular}}
\caption{Evaluation sets. $n$ is the md5-selected half of each 1{,}000-query
\hipporagtwo{} evaluation file (an earlier draft called these the
\hipporagtwo{} paper's own splits, which they are not). LastHop@$K$ counts a
query as covered iff the target passage's title is in the retrieved top-$K$.}
\label{tab:datasets}
\end{table}

\paragraph{Pipelines.}
We run \hipporagtwo{} via the official Python pipeline (\textsc{NV-Embed-v2}
encoder; default PPR step size); calibrated fusion as in \phasegraph{} (PIT
normalization followed by Boltzmann mixing at per-dataset $\alpha$); LLM
rerank via Nebius Llama-3.3-70B-Instruct at \$0.20/M prompt and \$0.60/M
completion tokens (the only paid service in the entire pipeline).
We re-use the cached \hipporagtwo{} top-10 and \phasegraph{} top-10 lists per
query from the head-to-head runs documented in our SOTA pipeline manifest;
both are deterministic given a fixed corpus and encoder.
The LLM rerank scores are cached in append-only JSONL caches per dataset
(\texttt{llm\_rerank\_cache\_*.jsonl}) so all numbers
in this paper are reproducible from cache with zero new LLM calls.
All gate evaluations are 5-fold cross-fitted on the test split itself with
out-of-fold predictions, so reported savings are out-of-sample by
construction; the threshold sweep is also performed on out-of-fold folds and
locked before reporting the test-split savings.

\subsection{Main result: rerank lift over \hipporagtwo{}}
\label{sec:exp:main}

Table~\ref{tab:main-lift} reports LastHop@$K$ for \hipporagtwo{} alone and
for the always-rerank pipeline on the corrected targets, with paired $W/L$
counts and exact two-sided sign-test $p$-values.
The reranker wins at every cutoff on 2Wiki and MuSiQue and at $K{=}1$ on
HotpotQA, by $+2.5$ to $+34.8$\,pp; HotpotQA at $K{=}5$ and $10$ is saturated
($\geq 0.95$ for both pipelines) and the two tie.
The MuSiQue row is the one that changes most from the earlier version: against
the terminal-hop passage the reranker gains $10.9$\,pp at $K{=}1$
($125/69$) where it previously appeared to lose $10$\,pp.
There is therefore no cell on which ``never rerank'' is the right policy for
coverage, and the two saturated HotpotQA cells are the only ones where it is
free.
\begin{table*}[t]
\centering
\small
\begin{tabular}{l c c c c l}
\toprule
Dataset & $K$ & \hipporagtwo{} & Always-rerank & $\Delta$ & Paired $W/L$ ($p$) \\
\midrule
2Wiki & 1 & 0.1385 & 0.4868 & $+34.8$\,pp & 221/50 ($p{=}8.0e-27$, rerank wins) \\
2Wiki & 5 & 0.6232 & 0.8004 & $+17.7$\,pp & 87/0 ($p{=}1.3e-26$, rerank wins) \\
2Wiki & 10 & 0.7067 & 0.8024 & $+9.6$\,pp & 47/0 ($p{=}1.4e-14$, rerank wins) \\
MuSiQue & 1 & 0.1868 & 0.2957 & $+10.9$\,pp & 125/69 ($p{=}7.1e-5$, rerank wins) \\
MuSiQue & 5 & 0.6868 & 0.7276 & $+4.1$\,pp & 42/21 ($p{=}0.011$, rerank wins) \\
MuSiQue & 10 & 0.7821 & 0.8074 & $+2.5$\,pp & 13/0 ($p{=}0.000$, rerank wins) \\
HotpotQA & 1 & 0.3397 & 0.5547 & $+21.5$\,pp & 207/95 ($p{=}1.0e-10$, rerank wins) \\
HotpotQA & 5 & 0.9482 & 0.9482 & $+0.0$\,pp & 10/10 ($p{=}1.000$, n.s.) \\
HotpotQA & 10 & 0.9712 & 0.9750 & $+0.4$\,pp & 2/0 ($p{=}0.500$, n.s.) \\
\bottomrule%
\end{tabular}
\caption{LastHop@$K$ for \hipporagtwo{} vs.\ the always-rerank pipeline on
the corrected targets (Section~\ref{sec:exp:setup}), all evaluation queries.
$W/L$ counts queries where only one of the two covers the target;
``n.s.''\ means the sign test does not reach $p<0.05$. Generated from
\texttt{paper14/repair/results/*\_nested.json}.}
\label{tab:main-lift}
\end{table*}

\subsection{Held-out gate results}
\label{sec:exp:gate}

Table~\ref{tab:gate-heldout} is the paper's main result: for each cell, the
policy selected inside the training folds and applied once to the held-out
folds (Section~\ref{sec:method:nested}), pooled over the five folds.
Averaged over the nine cells the gate skips $51.4\%$ of reranker calls at a
mean held-out coverage change of $-1.2$\,pp.
Four cells meet the pre-registered rule: 2Wiki $K{=}10$ ($36.5\%$ skipped,
$-0.81$\,pp), MuSiQue $K{=}10$ ($45.1\%$, $\pm 0.00$\,pp with no harmful or
beneficial skip at all), and the two saturated HotpotQA cells, where the gate
learned to skip nearly everything ($97.7\%$ and $99.6\%$) at
$-0.19$ and $+0.19$\,pp.
The other five cells buy $21$ to $54\%$ savings at $1.2$ to $3.3$\,pp: the
three high-lift $K{=}1$ cells skip $21$ to $51\%$ of calls and each harms
$44$ to $59$ queries while helping $27$ to $50$; 2Wiki $K{=}5$ skips only
$21\%$ and still harms six queries for no gain; MuSiQue $K{=}5$ skips $54\%$
at $-1.17$\,pp.
Harmful skips occur in eight of nine cells, $190$ in total against $136$
beneficial ones.
The inner selection preferred the distilled fallback $D$ on the 2Wiki cells,
where $D$ covers $4$ to $13$\,pp more than $H$ (Table~\ref{tab:baselines}),
and \hipporagtwo{} on MuSiQue and the HotpotQA $K{=}1$ and $5$ cells, where
the two are close; the choice was not stable across folds on five cells.
The earlier version's $73\%$ figure is not recoverable from any cell here:
its three highest cells were the oracle-fallback cells, and its MuSiQue
$K{=}1$ cell skipped every query against a target the reranker in fact
improves.
\begin{table*}[t]
\centering
\small
\resizebox{\textwidth}{!}{%
\begin{tabular}{l c c c c c c c c c}
\toprule
Dataset & $K$ & Fallback & Skip \% & Always-rerank & Gated & $\Delta$\,pp & Harmful & Beneficial & $\geq -1$\,pp \\
\midrule
2Wiki & 1 & D/H & 36.9 & 0.4868 & 0.4623 & $-2.44$ & 48 & 36 & $\times$ \\
2Wiki & 5 & D/H & 21.4 & 0.8004 & 0.7882 & $-1.22$ & 6 & 0 & $\times$ \\
2Wiki & 10 & D & 36.5 & 0.8024 & 0.7943 & $-0.81$ & 5 & 1 & \checkmark \\
MuSiQue & 1 & H & 50.6 & 0.2957 & 0.2782 & $-1.75$ & 59 & 50 & $\times$ \\
MuSiQue & 5 & H & 53.5 & 0.7276 & 0.7160 & $-1.17$ & 17 & 11 & $\times$ \\
MuSiQue & 10 & D/H & 45.1 & 0.8074 & 0.8074 & $+0.00$ & 0 & 0 & \checkmark \\
HotpotQA & 1 & D/H & 20.9 & 0.5547 & 0.5221 & $-3.26$ & 44 & 27 & $\times$ \\
HotpotQA & 5 & H & 97.7 & 0.9482 & 0.9463 & $-0.19$ & 10 & 9 & \checkmark \\
HotpotQA & 10 & D/H & 99.6 & 0.9750 & 0.9770 & $+0.19$ & 1 & 2 & \checkmark \\
\bottomrule%
\end{tabular}}
\caption{Held-out gate results, pooled over five outer folds. ``Fallback''
lists the fallback(s) the inner selection chose across folds ($H$:
\hipporagtwo{} top-$K$; $D$: distilled top-$K$). ``Harmful'' counts skipped
queries where the rerank would have covered the target and the fallback did
not; ``Beneficial'' the reverse. The last column applies the pre-registered
rule (held-out coverage within $1.0$\,pp of always-rerank).}
\label{tab:gate-heldout}
\end{table*}

\paragraph{Baselines.}
Table~\ref{tab:baselines} puts the learned gate next to the alternatives a
practitioner would try first, each selected with the same constraint inside
the same training folds: a one-feature gate on the \hipporagtwo{} top-1
margin; the best \emph{fixed} action per cell (always rerank, always $H$ or
always $D$); and a random gate that skips the same fraction of queries as
the learned gate, whose expected harm follows from the fallback's miss rate
alone.
The last column is the oracle skip rate that preserves every individual
query, $1-\Pr[\text{rerank hits}\wedge\text{fallback misses}]$, which is the
ceiling for a per-query-safe policy and is not a coverage bound.
The score-gap gate almost never finds a threshold that satisfies the
constraint inside the fold and ends up skipping $0$ to $2\%$ of queries,
except on saturated HotpotQA $K{=}5$ where a wide margin is a safe signal
($91\%$ skipped at $-0.2$\,pp).
The best fixed action is ``always rerank'' on every non-saturated cell; on
HotpotQA $K{=}5$ and $10$ the training folds chose a free action in four and
three of five folds, at $-1.0$ and $-0.2$\,pp, which the learned gate matches
with less loss.
The random gate at matched cost is the informative comparison: on the seven
non-saturated cells it loses $0.5$ to $11.1$\,pp where the learned gate loses
$0.0$ to $3.3$\,pp, so the gate's decisions concentrate the skips on queries
the fallback does cover.
The oracle column shows the ceiling for a policy that harms no query at all:
$55$ to $100\%$ with $H$ and $64$ to $100\%$ with $D$; the learned gate reaches
a quarter to two thirds of that ceiling on the non-saturated cells.
\begin{table*}[t]
\centering
\small
\resizebox{\textwidth}{!}{%
\begin{tabular}{l c c c c c c c c}
\toprule
 & & Learned gate & Score-gap gate & Best fixed & Random (matched) & Always $H$ & Always $D$ & Oracle skip \% \\
Dataset & $K$ & skip\,/\,$\Delta$ & skip\,/\,$\Delta$ & skip\,/\,$\Delta$ & skip\,/\,$\Delta$ & $\Delta$ & $\Delta$ & $H$\,/\,$D$ \\
\midrule
2Wiki & 1 & 36.9 / $-2.4$ & 0.2 / $-0.2$ & 0 / $+0.0$ & 36.9 / $-11.1$ & $-34.8$ & $-21.6$ & 55 / 69 \\
2Wiki & 5 & 21.4 / $-1.2$ & 0.0 / $+0.0$ & 0 / $+0.0$ & 21.4 / $-3.0$ & $-17.7$ & $-13.2$ & 82 / 86 \\
2Wiki & 10 & 36.5 / $-0.8$ & 0.2 / $+0.0$ & 0 / $+0.0$ & 36.5 / $-1.9$ & $-9.6$ & $-5.1$ & 90 / 94 \\
MuSiQue & 1 & 50.6 / $-1.8$ & 0.2 / $+0.0$ & 0 / $+0.0$ & 50.6 / $-5.5$ & $-10.9$ & $-11.5$ & 76 / 79 \\
MuSiQue & 5 & 53.5 / $-1.2$ & 2.1 / $-0.2$ & 0 / $+0.0$ & 53.5 / $-2.2$ & $-4.1$ & $-5.8$ & 92 / 91 \\
MuSiQue & 10 & 45.1 / $+0.0$ & 1.4 / $+0.0$ & 0 / $+0.0$ & 45.1 / $-0.5$ & $-2.5$ & $-1.2$ & 97 / 99 \\
HotpotQA & 1 & 20.9 / $-3.3$ & 0.0 / $+0.0$ & 0 / $+0.0$ & 20.9 / $-4.6$ & $-21.5$ & $-22.6$ & 60 / 64 \\
HotpotQA & 5 & 97.7 / $-0.2$ & 91.4 / $-0.2$ & 80 / $-1.0$ & 97.7 / $+0.0$ & $+0.0$ & $-1.0$ & 98 / 97 \\
HotpotQA & 10 & 99.6 / $+0.2$ & 1.9 / $-0.2$ & 60 / $-0.2$ & 99.6 / $+0.2$ & $-0.4$ & $+0.2$ & 100 / 100 \\
\bottomrule%
\end{tabular}}
\caption{Baselines under the same nested protocol. Each entry is held-out
skip rate (\%) and coverage change vs.\ always-rerank (pp). The random gate's
harm is an expectation at the learned gate's skip rate. ``Oracle skip''
is the per-query-preserving ceiling for each fallback.}
\label{tab:baselines}
\end{table*}

\paragraph{Optimism of in-fold selection.}
Table~\ref{tab:optimism} compares, per cell, the savings the inner selection
predicted for its chosen policy with the savings realised on the held-out
folds, and lists what was chosen in each fold.
The inner selection is only mildly optimistic about savings: the
nine-cell mean of its predictions is $52.4\%$ against $51.4\%$ realised, and
no cell is off by more than $3.2$\,pp.
Its optimism shows instead in the coverage constraint, which every selected
policy satisfied at margin zero inside the fold and five of nine violated
by $1.2$ to $3.3$\,pp outside it.
The per-fold predictions vary widely (MuSiQue $K{=}5$ from $39$ to $75\%$)
and the selected model and label change from fold to fold on most cells,
which is a sample-size statement: with $\approx 400$ training queries per
fold the constrained sweep over $48$ configurations and $51$ thresholds is
choosing among policies it cannot distinguish.
\begin{table}[h]
\centering
\small
\resizebox{\columnwidth}{!}{%
\begin{tabular}{l c c c c c l l}
\toprule
Dataset & $K$ & Inner & Held-out & Gap & Range & Model & Label \\
\midrule
2Wiki & 1 & 35.9 & 36.9 & $+0.9$ & 24--41 & lr/lr/lr/rf/lr & h/g/h/g/h \\
2Wiki & 5 & 22.1 & 21.4 & $-0.8$ & 13--29 & gbm/rf/lr/gbm/rf & g/g/g/g/g \\
2Wiki & 10 & 37.9 & 36.5 & $-1.4$ & 23--52 & rf/rf/lr/lr/rf & g/g/h/g/g \\
MuSiQue & 1 & 52.0 & 50.6 & $-1.5$ & 41--62 & lr/gbm/lr/lr/lr & g/g/g/g/g \\
MuSiQue & 5 & 56.7 & 53.5 & $-3.2$ & 39--75 & gbm/gbm/gbm/lr/gbm & g/g/g/h/g \\
MuSiQue & 10 & 47.3 & 45.1 & $-2.2$ & 33--70 & lr/lr/lr/lr/rf & g/g/g/g/g \\
HotpotQA & 1 & 22.1 & 20.9 & $-1.2$ & 15--33 & rf/rf/lr/lr/lr & g/g/g/g/g \\
HotpotQA & 5 & 98.4 & 97.7 & $-0.7$ & 92--100 & lr/lr/lr/lr/lr & h/g/h/h/h \\
HotpotQA & 10 & 99.1 & 99.6 & $+0.5$ & 97--100 & lr/gbm/gbm/lr/lr & h/g/g/h/g \\
\bottomrule%
\end{tabular}}
\caption{Inner-predicted vs.\ held-out savings (\%). ``Range'' spans the five
folds' inner predictions; ``Model'' and ``Label'' list the per-fold choices
(h: \textsc{rerank\_helps}, g: \textsc{fallback\_has\_gold}, a: always
rerank).}
\label{tab:optimism}
\end{table}

\subsection{Gating at a pre-specified loss budget}
\label{sec:exp:budget}

\begin{figure}[t]
\centering
\begin{tikzpicture}
\begin{axis}[width=0.98\columnwidth, height=0.72\columnwidth, xlabel={calibrated $P(\text{harm})$, bin mean}, ylabel={observed harmful-skip rate},
  xmin=0, xmax=1, ymin=0, ymax=1, grid=major, legend pos=north west, legend style={font=\scriptsize, draw=none, fill=none}, tick label style={font=\scriptsize}, label style={font=\small}]
\addplot[gray, dashed, domain=0:1, samples=2] {x};
\addlegendentry{perfect calibration}
\addplot[only marks, mark=*, blue, mark size=2pt] table[x=pred, y=obs] {figdata/reliability_2wiki.dat};
\addlegendentry{2Wiki}
\addplot[only marks, mark=square*, red!70!black, mark size=2pt] table[x=pred, y=obs] {figdata/reliability_musique.dat};
\addlegendentry{MuSiQue}
\addplot[only marks, mark=triangle*, green!50!black, mark size=2pt] table[x=pred, y=obs] {figdata/reliability_hotpot.dat};
\addlegendentry{HotpotQA}
\end{axis}
\end{tikzpicture}
\caption{Reliability of the Platt-calibrated harm probability on the pooled
held-out rows at the $1$\,pp budget: ten equal-width bins, pooled over $K$
within each dataset. Bins above $0.5$ hold few queries and are noisy; the
per-bin counts are in the ancillary data files.}
\label{fig:reliability}
\end{figure}
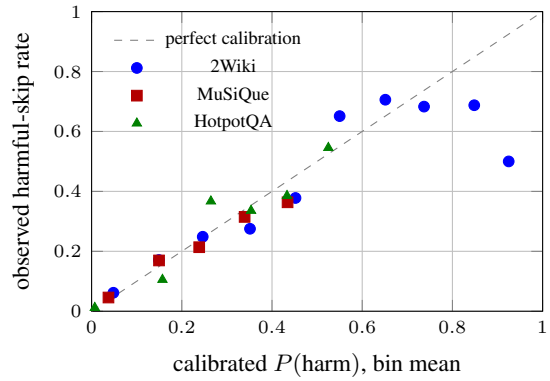

Table~\ref{tab:budget} scores the loss-budget rule of
Section~\ref{sec:method:budget} on the same held-out folds at the primary
budget of $1.0$\,pp expected harmful-skip rate, next to the coverage rule's
result for the cell.
Averaged over the nine cells the gate now skips $42.2\%$ of reranker calls
at $-0.77$\,pp, with $66$ harmful skips against $31$ beneficial ones (the
coverage rule: $51.4\%$, $-1.18$\,pp, $190$ against $136$), and six cells
are within $1$\,pp of always-rerank instead of four.
The savings move in both directions.
The three $K{=}1$ cells, which carried most of the harm under the coverage
rule, now skip $4$ to $8\%$ of calls and harm $4$ to $11$ queries instead of
$44$ to $59$; MuSiQue $K{=}10$, where the coverage rule found a $45\%$ skip
with no harm at all, is pushed by the budget to $95\%$ at $-2.14$\,pp with
$12$ harmful skips; the two saturated HotpotQA cells stay where they were.
The budget is met in expectation by construction, and in realisation in
three cells only: the pooled held-out harm is $1.45$\,pp on average against
a promised $0.83$\,pp, above the promise in six cells and by up to
$1.3$\,pp (MuSiQue $K{=}10$).
The gap is not miscalibration in aggregate.
After Platt scaling the expected calibration error of $\hat p(\text{harm})$
on the held-out rows is $0.001$ to $0.080$ per cell (mean $0.025$) against
$0.004$ to $0.150$ before (mean $0.094$), and the reliability diagram of
Figure~\ref{fig:reliability} sits close to the diagonal.
It is selection: the rule takes, among $24$ candidates, the one that
promises the most skips within the budget, and that candidate's promise is
biased downward exactly where it was chosen, in the low-probability tail of
its own calibrator; the in-fold optimism of Table~\ref{tab:optimism}
reappears as optimism about harm.
Table~\ref{tab:budget-sensitivity} gives the practical mapping: a $0.5$\,pp
budget realises a mean harm of $0.91$\,pp and a mean coverage change of
$-0.50$\,pp, with seven cells within $1$\,pp at $33.5\%$ savings; a $2$\,pp
budget returns to the coverage rule's savings ($50.4\%$) with $107$ harmful
skips and four cells within $1$\,pp.
A practitioner who wants a realised loss of about $1$\,pp should budget half
of it.
The calibrated probabilities are honest about the base rate and weak at
ranking: the area under the ROC curve of $\hat p(\text{harm})$ against the
realised harm label is $0.16$ to $0.70$ across cells (median $0.53$), below
$0.5$ on the two saturated HotpotQA cells with two and eight harmed queries.
The budget rule's savings therefore come from cells where harm is rare, not
from telling harmed queries apart, the same limit the oracle analysis of
Section~\ref{sec:analysis:oracle} points at.
\begin{table*}[t]
\centering
\small
\resizebox{\textwidth}{!}{%
\begin{tabular}{l c c c c c c c c c c c c}
\toprule
Dataset & $K$ & Cov.\ rule skip / $\Delta$ & Fallback & Skip \% & $\Delta$\,pp & Harmful & Beneficial & Expected harm & Realised harm & ECE cal. & ECE raw & $\geq -1$\,pp \\
\midrule
2Wiki & 1 & 36.9 / $-2.44$ & H & 6.9 & $+0.00$ & 11 & 11 & 1.03 & 2.24 & 0.080 & 0.067 & \checkmark \\
2Wiki & 5 & 21.4 / $-1.22$ & D/H & 19.8 & $-0.81$ & 4 & 0 & 0.90 & 0.81 & 0.021 & 0.139 & \checkmark \\
2Wiki & 10 & 36.5 / $-0.81$ & D & 39.7 & $-1.22$ & 8 & 2 & 1.03 & 1.63 & 0.014 & 0.135 & $\times$ \\
MuSiQue & 1 & 50.6 / $-1.75$ & D/H & 8.4 & $-0.97$ & 9 & 4 & 0.98 & 1.75 & 0.036 & 0.135 & \checkmark \\
MuSiQue & 5 & 53.5 / $-1.17$ & D/H & 14.8 & $-1.17$ & 8 & 2 & 0.67 & 1.56 & 0.008 & 0.102 & $\times$ \\
MuSiQue & 10 & 45.1 / $+0.00$ & D/H & 94.9 & $-2.14$ & 12 & 1 & 1.00 & 2.33 & 0.011 & 0.063 & $\times$ \\
HotpotQA & 1 & 20.9 / $-3.26$ & D & 4.2 & $-0.58$ & 4 & 1 & 0.80 & 0.77 & 0.043 & 0.150 & \checkmark \\
HotpotQA & 5 & 97.7 / $-0.19$ & H & 91.0 & $+0.00$ & 8 & 8 & 0.79 & 1.54 & 0.012 & 0.055 & \checkmark \\
HotpotQA & 10 & 99.6 / $+0.19$ & D/H & 99.8 & $+0.00$ & 2 & 2 & 0.23 & 0.38 & 0.001 & 0.004 & \checkmark \\
\bottomrule%
\end{tabular}}
\caption{Loss-budget gate at the primary budget of $1.0$\,pp expected
harmful-skip rate, pooled over five held-out folds, next to the coverage
rule's skip rate and coverage change for the same cell (Table~\ref{tab:gate-heldout}).
``Expected harm'' is the sum of the calibrated harm probabilities over the
skipped held-out queries divided by the number of queries, in pp; ``Realised
harm'' is the harmful-skip count on the same basis. ECE is the expected
calibration error of the harm probability on all held-out rows, after Platt
scaling (cal.) and before (raw).}
\label{tab:budget}
\end{table*}
\begin{table}[t]
\centering
\small
\resizebox{\columnwidth}{!}{%
\begin{tabular}{l c c c c c c c c}
\toprule
Rule & $\beta$ & Skip & $\Delta$ & Harm & Gain & $\geq -1$ & In budget & Worst \\
\midrule
coverage rule (v1) & -- & 51.4 & $-1.18$ & 190 & 136 & 4 & -- & -3.26 \\
loss budget & 0.5 & 33.5 & $-0.50$ & 42 & 19 & 7 & 3 & -1.75 \\
loss budget & 1.0 & 42.2 & $-0.77$ & 66 & 31 & 6 & 3 & -2.14 \\
loss budget & 2.0 & 50.4 & $-1.30$ & 107 & 48 & 4 & 3 & -3.26 \\
\bottomrule%
\end{tabular}}
\caption{Sensitivity to the budget $\beta$ (pp), macro-averaged over the
nine cells: skip rate (\%), coverage change (pp), harmful and beneficial
skip counts, cells within $1$\,pp of always-rerank, cells whose realised
harmful-skip rate stayed within the budget, and the largest held-out
coverage loss of any cell (pp).}
\label{tab:budget-sensitivity}
\end{table}

\subsection{Candidate-pool expansion}
\label{sec:exp:encoder}

We measure the LLM rerank's coverage when the candidate pool is extended from
two to five retrieval sources on 2Wiki (Table~\ref{tab:encoder}).
The 2-way pool uses only \textsc{NV-Embed-v2} (driving both \hipporagtwo{}'s
graph weights and \phasegraph{}'s dense channel, so two sources from one
encoder); the 3-way pool adds \textsc{BGE-large-en-v1.5}~\cite{xiao2023bge};
the 5-way pool further adds \textsc{Qwen3-Embedding-8B}~\cite{qwen3embed2025}
and \textsc{e5-mistral-7b-instruct}~\cite{wang2024e5mistral}.
This tests the reranker's robustness to pool expansion with the original
lists retained; it does not test the gate under an encoder replacement, which
remains untested (Section~\ref{sec:discussion}).
Each encoder contributes its own top-10 list to the merged set, which then
grows to a unique top-30 (capped) before the LLM rerank.

\begin{table*}[t]
\centering
\small
\begin{tabular}{l c c c c c c}
\toprule
Encoder pool & $K{=}1$ & $K{=}3$ & $K{=}5$ & $K{=}7$ & $K{=}10$ & Paired $p$ vs.\ 2-way \\
\midrule
2-way (\textsc{NV-Embed-v2} $+$ \phasegraph{}) & 0.487 & 0.782 & 0.800 & 0.800 & 0.802 & --- \\
3-way ($+$ \textsc{BGE-large}) & 0.511 & 0.780 & 0.804 & 0.804 & 0.807 & $[0.12, 1.0]$ \\
5-way ($+$ \textsc{Qwen3-8B}, \textsc{e5-mistral}) & 0.481 & 0.776 & 0.802 & 0.802 & 0.804 & $[0.64, 1.0]$ \\
\bottomrule
\end{tabular}
\caption{LLM-rerank coverage under candidate-pool expansion on 2Wiki,
$n\!=\!491$. Deltas vs.\ 2-way lie in $[-0.6, +2.4]$\,pp across
$K\!\in\!\{1,3,5,7,10\}$ (the $+2.4$\,pp is 3-way at $K\!=\!1$); no paired sign
test is significant, which is a failure to detect a difference, not evidence
of equivalence. Sources: \texttt{sota\_3way\_rerank\_results.json},
\texttt{sota\_llm\_rerank\_2wiki\_5way\_results.json}. MuSiQue and HotpotQA
were probed only with a single encoder swap and are not reported.}
\label{tab:encoder}
\end{table*}

The 5-way pool spends roughly twice the embedding-side compute of the 2-way
pool for changes of at most a few points in either direction, none
significant; the policy therefore stays on the 2-way pool.
The result says only that the listwise rerank's coverage is not sensitive to
adding candidates from other encoders while the original lists are kept; it
says nothing about how the gate's score-distribution features would behave if
\textsc{NV-Embed-v2} were replaced.

\subsection{Latency (component estimates only)}
\label{sec:exp:latency}

We did not measure end-to-end latency, and an earlier draft's per-query CDF
figure was a hand-drawn reconstruction that we have removed.
What was measured, over $n\!=\!20$ 2Wiki queries on an Apple-Silicon laptop
with \textsc{BGE-small} on MPS (\texttt{latency\_benchmark\_results.json}), is
the gate-feature embedding step alone: p50 $44.6$\,ms, p95 $262$\,ms, p99
$1{,}581$\,ms, the tail being \textsc{sentence-transformers}' first-call
warm-up.
The LLM call was not timed in that benchmark (zero calls logged); the
$\approx 1{,}000$\,ms figure used below is the Nebius default we assumed.
Adding the constant-time gate inference ($\approx 5$\,ms) and, for the
distilled fallback, per-passage scoring of $\leq 20$ candidates
($\approx 10$\,ms plus their embeddings), a skipped query costs of the order
of $400$\,ms at the tail versus of the order of $1{,}400$\,ms for an invoked
query, which carries the gate's overhead on top of the LLM call.
Because the p95 of a mixture is not the mixture of the p95s, we make no
claim about the end-to-end p95 of the gated policy; with a skip rate $s$ and
disjoint fast and slow paths, the policy's p95 sits inside the invoke
branch's distribution whenever $s<0.95$.
Measuring full traces for a fixed policy and workload, separating cold and
steady state, is left to the deployment (Section~\ref{sec:production}).

\subsection{Downstream QA F1}
\label{sec:exp:qa}

Table~\ref{tab:qa} reports answer-token F1 of a fixed Llama-3.3-70B answerer
over the top-5 passages of three \emph{always-on} pipelines; an earlier
draft labelled the always-rerank row as the gated pipeline, which it is not.
Table~\ref{tab:qa-replay} recombines the cached answers with the held-out
$K\!=\!5$ gate decisions where the caches allow it: the answers were
generated for \hipporagtwo{}'s and the reranker's top-5 only, so skipped
queries served from the distilled fallback have no cached answer and are
excluded, and the comparison is restricted to the covered queries.
On the covered queries the gated pipeline's answer F1 sits between
\hipporagtwo{} and always-rerank: $0.636$ against $0.643$ on 2Wiki ($416$ of
$491$ queries covered), $0.313$ against $0.317$ on MuSiQue and $0.685$
against $0.686$ on HotpotQA.
The gap to always-rerank is $0.1$ to $0.7$\,pp, smaller than the coverage
losses of Table~\ref{tab:gate-heldout}, which is consistent with the answerer
recovering some queries whose target is outside the top-5 and with the
earlier draft's observation that F1 saturates at the top of the list; the
comparison is nonetheless restricted to the covered queries and is not a
held-out measurement of the full gated pipeline.
\begin{table}[h]
\centering
\small
\resizebox{\columnwidth}{!}{%
\begin{tabular}{l c c c}
\toprule
Pipeline (always on) & 2Wiki & MuSiQue & HotpotQA \\
\midrule
\hipporagtwo{} top-5 & 0.576 & 0.307 & 0.686 \\
\phasegraph{} top-5  & 0.561 & 0.263 & 0.659 \\
Always-rerank top-5 & \textbf{0.659} & \textbf{0.317} & 0.686 \\
\bottomrule
\end{tabular}}
\caption{Answer F1 on top-5 passages, Llama-3.3-70B answerer at temperature
0, all evaluation queries, from the cached
\texttt{sota\_qa\_f1\_*\_results.json}. These rows never consult the gate.}
\label{tab:qa}
\end{table}
\begin{table}[h]
\centering
\small
\resizebox{\columnwidth}{!}{%
\begin{tabular}{l c c c c}
\toprule
Dataset & Covered & \hipporagtwo{} & Always-rerank & Gated \\
\midrule
2Wiki & 416/491 & 0.549 & 0.643 & 0.636 \\
MuSiQue & 514/514 & 0.307 & 0.317 & 0.313 \\
HotpotQA & 521/521 & 0.686 & 0.686 & 0.685 \\
\bottomrule%
\end{tabular}}
\caption{Gated answer F1 replayed from cached answers on the held-out
$K\!=\!5$ decisions, restricted to the covered queries (invoked queries and
$H$-fallback skips); the three columns are on the same queries.}
\label{tab:qa-replay}
\end{table}

\subsection{Held-out savings and coverage change per cell}
\label{sec:exp:heatmap}

Figure~\ref{fig:savings-heatmap} shows the held-out skip rate and the
coverage change of Table~\ref{tab:gate-heldout} as a grid, drawn directly
from the generated data file.
\begin{figure}[t]
\centering
\resizebox{\columnwidth}{!}{\begin{tikzpicture}[every node/.style={font=\scriptsize, anchor=center},
  cell/.style={draw, minimum width=1.9cm, minimum height=1.05cm, align=center, inner sep=2pt}]
\node at (-2.4, 1.3) {\textbf{Dataset}};
\node at (0, 1.3) {$K{=}1$}; \node at (2, 1.3) {$K{=}5$}; \node at (4, 1.3) {$K{=}10$};
\node at (-2.4, -0.00) {\textbf{2Wiki}};
\node[cell, fill=orange!45] at (0, -0.00) {36.9\%\\($-2.44$\,pp) };
\node[cell, fill=red!30] at (2, -0.00) {21.4\%\\($-1.22$\,pp) };
\node[cell, fill=orange!45] at (4, -0.00) {36.5\%\\($-0.81$\,pp) \checkmark};
\node at (-2.4, -1.10) {\textbf{MuSiQue}};
\node[cell, fill=green!30] at (0, -1.10) {50.6\%\\($-1.75$\,pp) };
\node[cell, fill=green!33] at (2, -1.10) {53.5\%\\($-1.17$\,pp) };
\node[cell, fill=orange!45] at (4, -1.10) {45.1\%\\($+0.00$\,pp) \checkmark};
\node at (-2.4, -2.20) {\textbf{HotpotQA}};
\node[cell, fill=red!30] at (0, -2.20) {20.9\%\\($-3.26$\,pp) };
\node[cell, fill=green!77] at (2, -2.20) {97.7\%\\($-0.19$\,pp) \checkmark};
\node[cell, fill=green!79] at (4, -2.20) {99.6\%\\($+0.19$\,pp) \checkmark};
\end{tikzpicture}}
\caption{Held-out skip rate (\%) with the held-out coverage change vs.\
always-rerank (pp, in parentheses) per (dataset, $K$) cell; \checkmark{}
marks the cells that meet the $1$\,pp rule. Colour encodes the skip rate
(red $<30\%$, orange $30$--$50\%$, green above). The figure is written by
\texttt{paper14/repair/make\_tables.py} from the nested-evaluation results,
as is \texttt{figdata/heldout\_savings.dat}.}
\label{fig:savings-heatmap}
\end{figure}

\subsection{Empirical coverage versus $K$}
\label{sec:exp:coverage}

Table~\ref{tab:coverage-k} reports the fraction of held-out queries whose
target is within the top $k$ of the delivered list, for the always-rerank
pipeline and for the $K\!=\!10$ gated policy, together with the smallest $k$
at which the empirical curve reaches $1-\alpha$.
These are descriptive curves on the pooled held-out folds, not conformal
guarantees: no separate calibration sample is held out and the gated lists
are not nested across cells.
The gated curves lie below always-rerank at small $k$ on every dataset
(for instance $0.448$ against $0.487$ at $k{=}1$ on 2Wiki), because the
$K{=}10$ policy defends coverage at $k{=}10$ only and the fallback it serves
is weaker at the head of the list; at $k{=}10$ the two curves meet within
the deltas of Table~\ref{tab:gate-heldout}.
The smallest $k$ reaching $75\%$ coverage grows from $3$ to $4$ on 2Wiki and
from $6$ to $7$ on MuSiQue under gating, and stays at $2$ on HotpotQA.
A policy that must defend several cutoffs at once needs a gate per cutoff or
a single gate trained on the strictest one; the cells here are independent.
\begin{table*}[t]
\centering
\small
\resizebox{\textwidth}{!}{%
\begin{tabular}{l l c c c c c c c c c}
\toprule
 & & \multicolumn{5}{c}{coverage at $k$} & \multicolumn{4}{c}{smallest $k$ reaching $1-\alpha$} \\
Dataset & Pipeline & 1 & 3 & 5 & 7 & 10 & $\alpha{=}.25$ & $.20$ & $.10$ & $.05$ \\
\midrule
2Wiki & always-rerank & 0.487 & 0.782 & 0.800 & 0.800 & 0.802 & 3 & 5 & -- & -- \\
2Wiki & gated ($K{=}10$ policy) & 0.448 & 0.729 & 0.770 & 0.780 & 0.794 & 4 & -- & -- & -- \\
MuSiQue & always-rerank & 0.296 & 0.613 & 0.728 & 0.774 & 0.807 & 6 & 9 & -- & -- \\
MuSiQue & gated ($K{=}10$ policy) & 0.257 & 0.597 & 0.718 & 0.765 & 0.807 & 7 & 10 & -- & -- \\
HotpotQA & always-rerank & 0.555 & 0.902 & 0.948 & 0.969 & 0.975 & 2 & 2 & 3 & 6 \\
HotpotQA & gated ($K{=}10$ policy) & 0.338 & 0.856 & 0.939 & 0.962 & 0.977 & 2 & 3 & 4 & 6 \\
\bottomrule%
\end{tabular}}
\caption{Empirical coverage versus $k$ on the held-out folds. ``--'' means no
$k\leq 10$ reaches the level. The gated rows use the $K\!=\!10$ cell's
decisions.}
\label{tab:coverage-k}
\end{table*}

\subsection{Cost}
\label{sec:exp:cost}

Table~\ref{tab:costs} converts the held-out decisions into LLM-rerank cost
per 1{,}000 queries at the historical Nebius prices used throughout
(\$0.20 per million prompt tokens, \$0.60 per million completion tokens).
For MuSiQue and HotpotQA every rerank call's token counts are cached, so the
gated cost sums the invoked queries' actual tokens; the 2Wiki cache has no
per-call counts and uses the archived per-call average, so its cost saving
equals its skip rate by construction.
The figures are marginal reranker cost only; \hipporagtwo{}'s own online
LLM triple filtering and the answerer are outside them.
On the two datasets with measured tokens the cost saving is $1$ to $4$\,pp
below the skip rate (MuSiQue $46.6$ to $52.4\%$ saved against $45.1$ to
$53.5\%$ skipped), because the queries the gate sends to the reranker carry
slightly longer prompts than the ones it skips.
The nine-cell mean cost saving is $50.5\%$, and the marginal reranker cost
falls from \$0.22--0.29 to \$0.10--0.23 per 1{,}000 queries on the
non-saturated cells and to under a cent on the two saturated ones.
\begin{table}[h]
\centering
\small
\resizebox{\columnwidth}{!}{%
\begin{tabular}{l c c c c c l}
\toprule
Dataset & $K$ & Always \$/1k & Gated \$/1k & Cost saved \% & Skip \% & Tokens \\
\midrule
2Wiki & 1 & 0.287 & 0.181 & 36.9 & 36.9 & per-call average \\
2Wiki & 5 & 0.287 & 0.225 & 21.4 & 21.4 & per-call average \\
2Wiki & 10 & 0.287 & 0.182 & 36.5 & 36.5 & per-call average \\
MuSiQue & 1 & 0.218 & 0.116 & 46.6 & 50.6 & measured \\
MuSiQue & 5 & 0.218 & 0.103 & 52.4 & 53.5 & measured \\
MuSiQue & 10 & 0.218 & 0.125 & 42.7 & 45.1 & measured \\
HotpotQA & 1 & 0.292 & 0.230 & 21.1 & 20.9 & measured \\
HotpotQA & 5 & 0.292 & 0.007 & 97.7 & 97.7 & measured \\
HotpotQA & 10 & 0.292 & 0.001 & 99.6 & 99.6 & measured \\
\bottomrule%
\end{tabular}}
\caption{Marginal LLM-rerank cost per 1{,}000 queries under the held-out
decisions. ``Tokens'' says whether per-query counts were measured or the
per-call average was used.}
\label{tab:costs}
\end{table}

\section{Analysis}
\label{sec:analysis}

\subsection{What an oracle could skip}
\label{sec:analysis:oracle}

An earlier draft reported a ``three-tier oracle'' as the per-query maximum of
\hipporagtwo{}, distillation and rerank coverage and treated its coverage as
a ceiling on the gate's savings.
Coverage is not a bound on savings: if the fallback covers exactly the
queries the reranker covers, every call can be skipped at no loss.
The relevant ceiling for a policy that must not harm any individual query is
the per-query-preserving oracle skip rate
$1-\Pr[\text{rerank hits}\wedge\text{fallback misses}]$, computed for each
fallback in the last column of Table~\ref{tab:baselines}.
On the non-saturated cells that ceiling is $55$ to $97\%$ with $H$ and
$64$ to $99\%$ with $D$, against realised gate skip rates of $21$ to $54\%$.
Under an aggregate constraint, losses may be offset by the queries the
fallback covers and the reranker does not, which is why
Table~\ref{tab:gate-heldout} reports the two counts separately.

\section{Production Deployment}
\label{sec:production}

A gate service is live on Railway and exposes two endpoints:
\texttt{GET /paper14/status} returns the loaded per-cell policy and the
\textsc{BGE-small} state, and \texttt{POST /paper14/gate\_predict} takes
$\{q, H(q), V(q), \text{dataset}, K\}$ and returns \texttt{invoke},
\texttt{meta} and the fallback top-$K$.
Two of the checks an earlier draft reported for it do not mean what that
draft said.
The ``paraphrase OOD test'' sent each of $450$ (query, dataset, $K$) triples
twice, once with the original question and once with a Llama-3.3-70B
paraphrase, while holding the retrieval lists fixed, and counted agreement of
the gate decision: $434/450$ agreed ($96.4\%$).
That is a serving and input-perturbation check of the gate's features; it
does not re-run retrieval and it measures no answer quality, and cells whose
policy skips or invokes almost every query agree trivially.
The ``end-to-end test-time audit'' evaluated the deployed classifiers on the
very queries they were fitted on and reported skip rate plus the negative
part of the net coverage change as a ``safe-skip rate''; neither is a
held-out measurement (Section~\ref{sec:method:nested} is).
The policy file the earlier draft deployed was fitted on all evaluation
queries, used the oracle fallback on three cells and the wrong MuSiQue
target on three others; it is withdrawn.
The repository now ships a policy retrained under the loss-budget protocol
of Section~\ref{sec:method:budget} on all queries of each cell
(\texttt{paper14/repair/train\_production\_policy.py}): the harm label,
five-fold out-of-fold probabilities, Platt calibration and the largest skip
set within the $1$\,pp budget, refitted once and shipped with its Platt pair
and threshold.
A cell is active only if its pooled held-out coverage change under the same
protocol (Table~\ref{tab:budget}) is within the pre-registered $1$\,pp: the
six cells that meet it (2Wiki $K{=}1$ and $5$, MuSiQue $K{=}1$, HotpotQA
$K{=}1$, $5$ and $10$) ship with the gate, the other three (2Wiki $K{=}10$,
MuSiQue $K{=}5$ and $10$) ship as always-invoke, and the service records
which applied in each decision's metadata.
Two caveats apply to any number quoted for the live service.
The serving path feeds the gate PortMem's own graph and vector channels, not
the official \hipporagtwo{} lists the evaluation used, so the held-out
numbers are the protocol's estimate for the policy, not a measurement of the
deployed one; and the harm counts of Table~\ref{tab:budget}, with their
realised-versus-expected gap, are the numbers a rollout decision should be
made on.

\section{Discussion and Limitations}
\label{sec:discussion}

\paragraph{Cross-dataset generalisation is untested.}
\system{} trains one gate per (dataset, $K$) cell and selects everything
inside that dataset's folds; transfer to a new dataset is not measured.
The per-fold choices of Table~\ref{tab:optimism} suggest that per-dataset
calibration is unavoidable: the label, model and fallback change from fold
to fold within a cell, tracking the cell's positive-class density rather
than any stable rule.
A reasonable proxy would hold out one of the three datasets, train the
other two with shared hyperparameters and report transfer savings; we have
not run it and flag it as the most important follow-up for any adopter.

\paragraph{QA F1 saturation at the top of the list.}
On HotpotQA the final-hop passage is in the top-5 of every pipeline for
about $95\%$ of queries, and the answerer's F1 does not move with the
reranker ($0.686$ either way, Table~\ref{tab:qa}); on 2Wiki the reranker's
$+17.7$\,pp at $K{=}5$ becomes $+8.3$\,pp of F1.
The gated replay of Table~\ref{tab:qa-replay} sits within $0.7$\,pp of
always-rerank on the covered queries, less than the coverage losses, which
suggests the answerer tolerates some of the fallback's misses; that replay
covers only the queries whose answers were cached and is not a full held-out
measurement of gated answer quality.

\paragraph{Candidate-pool expansion was measured on 2Wiki only.}
Table~\ref{tab:encoder} shows the reranker's coverage when candidates from
three further encoders are added to the pool on 2Wiki, with the original
lists kept: changes within $[-0.6, +2.4]$\,pp, none significant, which is a
failure to detect a difference rather than evidence of invariance.
Nothing in this paper tests the gate itself under an encoder replacement;
its features are score statistics of \textsc{NV-Embed-v2}-driven lists, and
a different encoder's score distribution would require retraining and
re-evaluation.
MuSiQue and HotpotQA were probed only with a single encoder swap and are not
reported.
\paragraph{MuSiQue and HotpotQA targets.}
An earlier draft scored MuSiQue against the last supporting paragraph in the
dataset's shuffled paragraph order, which is the terminal-hop passage for
only $45\%$ of test queries; under that target the reranker appeared to lose
$10$\,pp at $K\!=\!1$, and a whole analysis section explained why.
Against the terminal decomposition passage the reranker wins at every $K$
(Table~\ref{tab:main-lift}), and that section is gone.
HotpotQA's target is the unique answer-bearing supporting passage where one
exists ($398/521$ test queries) and the last listed supporting fact otherwise
(comparison questions, and bridge questions whose answer string appears in
both passages); 2Wiki's is the last listed supporting fact, which is the
answer-bearing passage for compositional and inference questions
($253/254$) and one of the two compared entities for comparison questions.
All three are title-level metrics on the HippoRAG~2 side; titles repeat in
the corpora (on MuSiQue $189/514$ target titles are shared), which inflates
coverage by roughly $2$\,pp where we could measure it on passage-identified
lists.

\paragraph{Calibrated in aggregate, optimistic after selection.}
The Platt-calibrated harm probabilities of Section~\ref{sec:exp:budget}
have small expected calibration error on held-out rows, yet the skip sets
chosen by their own expected harm realise $0.6$\,pp more harm than promised
on average.
Choosing the most aggressive of $24$ candidates by a quantity estimated on
the same inner folds is a winner's curse, and a budget stated to a user
should be discounted for it; a split of the training fold between
calibration and selection, or a conformal bound on the harm count, would
trade savings for a promise that holds, and we have run neither.
The probabilities also rank harm poorly (AUC $0.16$ to $0.70$), so the
achievable savings at any budget are set by each cell's harm base rate
rather than by the classifier.

\paragraph{Gate trained per-cell, not per-query class.}
The gate treats all queries in a cell uniformly.
Routing work such as Adaptive-RAG~\cite{jeong2024adaptive} and
Self-RAG~\cite{asai2024self} benefits from explicit query-type
classification; an upstream query-type classifier choosing the fallback or
the threshold per query is a natural extension, and the oracle column of
Table~\ref{tab:baselines} bounds what any such refinement could add.

\paragraph{Distillation teacher is the same LLM as the rerank.}
The distilled fallback regresses on the cached Llama-3.3-70B scores, so it
inherits the reranker's errors and cannot exceed it; where the distilled
top-$K$ covers more than \hipporagtwo{}'s (2Wiki, by $4$ to $13$\,pp) the
gate chose it, where it does not the gate fell back on \hipporagtwo{}.
A different teacher would move both the fallback and the reranker, and with
them the whole trade-off; we have not tried one.

\section{Conclusion}
\label{sec:conclusion}

LLM rerankers are a real accuracy boost for graph-augmented multi-hop
retrieval and a real cost on every query.
We asked whether a small gate over pre-LLM features can tell, per query,
when the reranker is not needed, and we answered it under a protocol in
which nothing the gate needs is chosen on the queries it is scored on.
The answer is a qualified yes: the gate skips half of the reranker calls on
average, four of nine cells are non-inferior at $1$\,pp, the decisions are
far better than random skipping at the same cost on every high-lift cell,
and the cost saving tracks the skip rate.
It is also a plain no to the claim the earlier version made: the gate is not
lossless, harmful skips occur in eight of nine cells, and the five cells with
the largest reranker lift pay $1.2$ to $3.3$\,pp for $21$ to $54\%$ savings.
Whether that trade is worth taking is a deployment decision that
Table~\ref{tab:gate-heldout} now gives the numbers for.
Setting the threshold from a stated expected-harm budget over calibrated
probabilities (Section~\ref{sec:exp:budget}) makes the trade explicit:
$42\%$ of calls skipped at $-0.8$\,pp for a $1$\,pp budget, six cells within
$1$\,pp, and a realised harm about half again what was promised, which a
user should budget for.
The natural follow-ups are more training queries per cell, since the
in-fold selection is choosing among policies it cannot distinguish; a
selection-aware or conformal bound on the harm count; a gate that defends
several cutoffs at once; passage-level rather than title-level targets; and
a test of the gate under an encoder replacement, which this paper does not
contain.

\bibliography{references}

\begin{thebibliography}{34}
\providecommand{\natexlab}[1]{#1}

\bibitem[{Angelopoulos and Bates(2021)}]{angelopoulos2021gentle}
Anastasios~N. Angelopoulos and Stephen Bates. 2021.
\newblock A gentle introduction to conformal prediction and distribution-free
  uncertainty quantification.
\newblock \emph{arXiv preprint arXiv:2107.07511}.

\bibitem[{Asai et~al.(2024)Asai, Wu, Wang, Sil, and Hajishirzi}]{asai2024self}
Akari Asai, Zeqiu Wu, Yizhong Wang, Avirup Sil, and Hannaneh Hajishirzi. 2024.
\newblock {Self-RAG}: Learning to retrieve, generate, and critique through
  self-reflection.
\newblock In \emph{Proceedings of ICLR}.

\bibitem[{Bacellar(2026{\natexlab{a}})}]{bacellar2026abstain}
Andre Bacellar. 2026{\natexlab{a}}.
\newblock Predictable failure in multi-hop retrieval: Score-distributional
  confidence scoring and abstention.
\newblock \emph{arXiv preprint}.
\newblock Submitted September 2026.

\bibitem[{Bacellar(2026{\natexlab{b}})}]{bacellar2026regime}
Andre Bacellar. 2026{\natexlab{b}}.
\newblock Regime-conditional retrieval: Theory and a transferable router for
  two-hop {QA}.
\newblock \emph{arXiv preprint arXiv:2604.09019}.

\bibitem[{Bansal and Agarwal(2026)}]{bansal2026ragrouter}
Prakhar Bansal and Shivangi Agarwal. 2026.
\newblock Lightweight query routing for adaptive {RAG}: A baseline study on
  {RAGRouter-Bench}.
\newblock \emph{arXiv preprint arXiv:2604.03455}.

\bibitem[{Cormack et~al.(2009)Cormack, Clarke, and
  Buettcher}]{cormack2009reciprocal}
Gordon~V. Cormack, Charles~L.A. Clarke, and Stefan Buettcher. 2009.
\newblock Reciprocal rank fusion outperforms {Condorcet} and individual rank
  learning methods.
\newblock In \emph{Proceedings of SIGIR}, pages 758--759.

\bibitem[{Dabod et~al.(2026)Dabod, Cohen, and
  Stanovsky}]{dabod2026rerankinghurts}
Orian Dabod, Amir D.~N. Cohen, and Gabriel Stanovsky. 2026.
\newblock When reranking hurts: Uncertainty-based gating for few-shot
  reranking.
\newblock \emph{arXiv preprint arXiv:2606.31087}.

\bibitem[{Dubey et~al.(2024)Dubey, Jauhri, Ghosh, and {Meta AI}}]{llama3_2024}
Abhimanyu Dubey, Abhinav Jauhri, Abhinav Ghosh, and {Meta AI}. 2024.
\newblock The {Llama} 3 herd of models.
\newblock \emph{arXiv preprint arXiv:2407.21783}.
\newblock We use Llama-3.3-70B-Instruct as the reranker LLM.

\bibitem[{Genc et~al.(2026)Genc, Korukluoglu, and Allan}]{genc2026adaptiverr}
Ata~Cinar Genc, Emir~Kaan Korukluoglu, and James Allan. 2026.
\newblock Adaptive re-ranking.
\newblock \emph{arXiv preprint arXiv:2606.25249}.

\bibitem[{Gonzalez-Gamella et~al.(2026)Gonzalez-Gamella, Sanchez-Santolaya,
  Moraleda-Moreno, and Chaquet-Ulldemolins}]{gonzalez2026ards}
Carlos Gonzalez-Gamella, Daniel Sanchez-Santolaya, Francisco Moraleda-Moreno,
  and Jacobo Chaquet-Ulldemolins. 2026.
\newblock \href {https://doi.org/10.5220/0014803000004018} {When to rerank?
  {Adaptive} reranking decisions for efficient information retrieval}.
\newblock In \emph{Proceedings of the 28th International Conference on
  Enterprise Information Systems (ICEIS), Volume 1}, pages 545--556.

\bibitem[{Guo et~al.(2017)Guo, Pleiss, Sun, and
  Weinberger}]{guo2017calibration}
Chuan Guo, Geoff Pleiss, Yu~Sun, and Kilian~Q. Weinberger. 2017.
\newblock On calibration of modern neural networks.
\newblock In \emph{Proceedings of ICML}, pages 1321--1330.

\bibitem[{Guti{\'e}rrez et~al.(2024)Guti{\'e}rrez, Shu, Gu, Yasunaga, and
  Su}]{gutierrez2024hipporag}
Bernal~Jim{\'e}nez Guti{\'e}rrez, Yiheng Shu, Yu~Gu, Michihiro Yasunaga, and
  Yu~Su. 2024.
\newblock {HippoRAG}: Neurobiologically inspired long-term memory for large
  language models.
\newblock In \emph{Proceedings of NeurIPS}.

\bibitem[{Guti{\'e}rrez et~al.(2025)Guti{\'e}rrez, Shu, Qi, Zhou, and
  Su}]{gutierrez2025hipporag2}
Bernal~Jim{\'e}nez Guti{\'e}rrez, Yiheng Shu, Weijian Qi, Sizhe Zhou, and
  Yu~Su. 2025.
\newblock From {RAG} to memory: Non-parametric continual learning for large
  language models.
\newblock \emph{arXiv preprint arXiv:2502.14802}.

\bibitem[{Ho et~al.(2020)Ho, Duong~Nguyen, Sugawara, and Aizawa}]{ho2020wiki2}
Xanh Ho, Anh-Khoa Duong~Nguyen, Saku Sugawara, and Akiko Aizawa. 2020.
\newblock Constructing a multi-hop {QA} dataset for comprehensive evaluation of
  reasoning steps.
\newblock In \emph{Proceedings of COLING}, pages 6609--6625.

\bibitem[{Jeong et~al.(2024)Jeong, Baek, Cho, Hwang, and
  Park}]{jeong2024adaptive}
Soyeong Jeong, Jinheon Baek, Sukmin Cho, Sung~Ju Hwang, and Jong~C. Park. 2024.
\newblock {Adaptive-RAG}: Learning to adapt retrieval-augmented large language
  models through question complexity.
\newblock In \emph{Proceedings of NAACL}.

\bibitem[{Jia et~al.(2026)Jia, Ye, Jia, Qian et~al.}]{jia2026balancerag}
Zijun Jia, Yuanchang Ye, Sen Jia, Yiyao Qian, and 1 others. 2026.
\newblock {BalanceRAG}: Joint risk calibration for cascaded retrieval-augmented
  generation.
\newblock \emph{arXiv preprint arXiv:2605.20084}.

\bibitem[{Khattab and Zaharia(2020)}]{khattab2020colbert}
Omar Khattab and Matei Zaharia. 2020.
\newblock {ColBERT}: Efficient and effective passage search via contextualized
  late interaction over {BERT}.
\newblock In \emph{Proceedings of SIGIR}, pages 39--48.

\bibitem[{Lee et~al.(2024)Lee, Roy, Xu, Raiman, Shoeybi, Catanzaro, and
  Ping}]{nvembed2024}
Chankyu Lee, Rajarshi Roy, Mengyao Xu, Jonathan Raiman, Mohammad Shoeybi, Bryan
  Catanzaro, and Wei Ping. 2024.
\newblock {NV-Embed}: Improved techniques for training {LLMs} as generalist
  embedding models.
\newblock \emph{arXiv preprint arXiv:2405.17428}.

\bibitem[{Li et~al.(2024)Li, Li, Zhang, Mei, and Bendersky}]{wang2024selfroute}
Zhuowan Li, Cheng Li, Mingyang Zhang, Qiaozhu Mei, and Michael Bendersky. 2024.
\newblock Retrieval augmented generation or long-context {LLMs}? {A}
  comprehensive study and hybrid approach.
\newblock \emph{arXiv preprint arXiv:2407.16833}.
\newblock Introduces the Self-Route method.

\bibitem[{Miranda(2026{\natexlab{a}})}]{bacellar2026bridgerag}
Andre Miranda. 2026{\natexlab{a}}.
\newblock {BridgeRAG}: Bridge-conditioned retrieval for multi-hop {QA}.
\newblock \emph{arXiv preprint arXiv:2604.03384}.

\bibitem[{Miranda(2026{\natexlab{b}})}]{phasegraph2026}
Andre Miranda. 2026{\natexlab{b}}.
\newblock Calibrated fusion for heterogeneous graph-vector retrieval in
  multi-hop {QA}.
\newblock \emph{arXiv preprint arXiv:2603.28886}.

\bibitem[{Mishra(2026)}]{mishra2026costaware}
Sanjay Mishra. 2026.
\newblock Cost-aware query routing in {RAG}: Empirical analysis of retrieval
  depth tradeoffs.
\newblock \emph{arXiv preprint arXiv:2606.02581}.

\bibitem[{Nogueira and Cho(2019)}]{nogueira2019passage}
Rodrigo Nogueira and Kyunghyun Cho. 2019.
\newblock Passage re-ranking with {BERT}.
\newblock In \emph{arXiv preprint arXiv:1901.04085}.

\bibitem[{Platt(1999)}]{platt1999}
John~C. Platt. 1999.
\newblock Probabilistic outputs for support vector machines and comparisons to
  regularized likelihood methods.
\newblock In \emph{Advances in Large Margin Classifiers}, pages 61--74. MIT
  Press.

\bibitem[{Pradeep et~al.(2023)Pradeep, Sharifymoghaddam, and
  Lin}]{pradeep2023rankzephyr}
Ronak Pradeep, Sahel Sharifymoghaddam, and Jimmy Lin. 2023.
\newblock {RankZephyr}: Effective and robust zero-shot listwise reranking is a
  breeze!
\newblock \emph{arXiv preprint arXiv:2312.02724}.

\bibitem[{Sun et~al.(2023)Sun, Yan, Ma, Wang, Ren, Chen, Yin, and
  Ren}]{sun2023rankgpt}
Weiwei Sun, Lingyong Yan, Xinyu Ma, Shuaiqiang Wang, Pengjie Ren, Zhumin Chen,
  Dawei Yin, and Zhaochun Ren. 2023.
\newblock Is {ChatGPT} good at search? investigating large language models as
  re-ranking agents.
\newblock In \emph{Proceedings of EMNLP}.

\bibitem[{Trivedi et~al.(2022)Trivedi, Balasubramanian, Khot, and
  Sabharwal}]{trivedi2022musique}
Harsh Trivedi, Niranjan Balasubramanian, Tushar Khot, and Ashish Sabharwal.
  2022.
\newblock {MuSiQue}: Multihop questions via single-hop question composition.
\newblock \emph{Transactions of the Association for Computational Linguistics},
  10:539--554.

\bibitem[{Vovk et~al.(2005)Vovk, Gammerman, and Shafer}]{vovk2005algorithmic}
Vladimir Vovk, Alex Gammerman, and Glenn Shafer. 2005.
\newblock \emph{Algorithmic Learning in a Random World}.
\newblock Springer.

\bibitem[{Wang and Han(2025)}]{proprag2025}
Jingjin Wang and Jiawei Han. 2025.
\newblock {PropRAG}: Guiding retrieval with beam search over proposition paths.
\newblock In \emph{Proceedings of EMNLP}, pages 6212--6227.

\bibitem[{Wang et~al.(2024)Wang, Yang, Huang, Yang, Majumder, and
  Wei}]{wang2024e5mistral}
Liang Wang, Nan Yang, Xiaolong Huang, Linjun Yang, Rangan Majumder, and Furu
  Wei. 2024.
\newblock Improving text embeddings with large language models.
\newblock \emph{arXiv preprint arXiv:2401.00368}.
\newblock The paper behind the e5-mistral-7b-instruct model.

\bibitem[{Wang et~al.(2025)Wang, Wei, and Ling}]{wang2025retrievaldecision}
Yufeng Wang, Lu~Wei, and Haibin Ling. 2025.
\newblock Retrieval as a decision: Training-free adaptive gating for efficient
  {RAG}.
\newblock \emph{arXiv preprint arXiv:2511.09803}.

\bibitem[{Xiao et~al.(2023)Xiao, Liu, Zhang, and Muennighoff}]{xiao2023bge}
Shitao Xiao, Zheng Liu, Peitian Zhang, and Niklas Muennighoff. 2023.
\newblock {C-Pack}: Packaged resources to advance general chinese embedding.
\newblock \emph{arXiv preprint arXiv:2309.07597}.
\newblock Source of \textsc{BGE-small-en-v1.5} and \textsc{BGE-large-en-v1.5}.

\bibitem[{Yang et~al.(2018)Yang, Qi, Zhang, Bengio, Cohen, Salakhutdinov, and
  Manning}]{yang2018hotpotqa}
Zhilin Yang, Peng Qi, Saizheng Zhang, Yoshua Bengio, William~W. Cohen, Ruslan
  Salakhutdinov, and Christopher~D. Manning. 2018.
\newblock {HotpotQA}: A dataset for diverse, explainable multi-hop question
  answering.
\newblock In \emph{Proceedings of EMNLP}, pages 2369--2380.

\bibitem[{Zhang et~al.(2025)Zhang, Li, Long, Zhang et~al.}]{qwen3embed2025}
Yanzhao Zhang, Mingxin Li, Dingkun Long, Xin Zhang, and 1 others. 2025.
\newblock {Qwen3 Embedding}: Advancing text embedding and reranking through
  foundation models.
\newblock \emph{arXiv preprint arXiv:2506.05176}.

\end{thebibliography}

\end{document}